\documentclass{aa}
\usepackage{astron}
\usepackage{graphics}
\usepackage{psfig}
\usepackage{times}
\def\la{\raise.5ex\hbox{$<$}\kern-.8em\lower 1mm\hbox{$\sim$}}
\def\ma{\raise.5ex\hbox{$>$}\kern-.8em\lower 1mm\hbox{$\sim$}}

\def\kms{$\rm km\, s^{-1}$}
\def\cm3{$\rm cm^{-3}$}
\def\Ts{$T_{\rm *}$}
\def\Vs{$V_{\rm s}$}
\def\n0{$n_{\rm 0}$}
\def\B0{$B_{\rm 0}$}
\def\ne{$n_{\rm e}$}
\def\Te{$T_{\rm e}$}

\def\erg{$\rm erg\, cm^{-2}\, s^{-1}$}
\def\ergs{$\rm erg\, s^{-1}$}
\def\hii{H{\sc ii}}

\def\oi{O\,{\sc i}}
\def\oiii{O\,{\sc iii}}
\def\oiv{O\,{\sc iv}}
\def\neii{Ne\,{\sc ii}}
\def\neiii{Ne\,{\sc iii}}
\def\siii{S\,{\sc iii}}
\def\slii{Si\,{\sc ii}}
\def\cii{C\,{\sc ii}}

\def\Cii{[C\,{\sc ii}]}
\def\Siii{[S\,{\sc iii}]}
\def\Oiii{[O\,{\sc iii}]}
\def\Oi{[O\,{\sc i}]}
\def\Slii{[Si\,{\sc ii}]}
\def\Oiv{[O\,{\sc iv}]}
\def\Neii{[Ne\,{\sc ii}]}
\def\Neiii{[Ne\,{\sc iii}]}
\begin{document}

   \thesaurus{(11.09.1; 11.09.4; 11.19.3; 13.09.1; 02.12.1; 02.19.1)} 

\title{Infrared line ratios revealing starburst conditions in galaxies}

\author{
Sueli M.\ Viegas\inst{1}, 
Marcella Contini\inst{1,2} 
\and Thierry Contini\inst{2,3}
}
   
\offprints{T.\ Contini, tcontini@eso.org}

   \institute{
Instituto Astron\^omico e Geof\' \i sico, USP, Av. Miguel Stefano 4200, 
04301-904 S\~ao Paulo, Brazil
\and
School of Physics \& Astronomy, Tel Aviv University, 
69978 Tel Aviv, Israel
\and 
European Southern Observatory, 
Karl-Schwarzschild-Str. 2, D-85748 Garching bei M\"unchen, Germany
}
   \date{Received 29 October 1998; accepted 16 April 1999}

\authorrunning{S.M.\ Viegas, M.\ Contini \& T.\ Contini}
\titlerunning{IR line ratios revealing starburst conditions in
galaxies}
   \maketitle

\begin{abstract}
The physical conditions in  typical starburst galaxies are investigated
through 
critical infrared (IR) line ratios, as previously suggested by 
Lutz et al.\ \cite*{LUTZetal98}.
The calculations by a composite model which consistently accounts for
the coupled effect of shock and photoionization by hot stars
definitely fit the observed line ratios of single objects and
explain the observed  relation between [\oiv]/([\neii]+0.44[\neiii]) 
and [\neiii]/[\neii].
The shock velocity and the gas density are the critical parameters.

Most of the shocks are produced in low density-velocity (\n0 $\sim$ 100 \cm3\ 
and \Vs\ $\sim 50 - 100$ \kms) clouds which represent the bulk of 
the ionized gas in starburst galaxies. 
However, though they are by many orders less numerous, high-velocity 
($\sim 400 - 600$ \kms) shocks in dense ($\sim 500 - 800$ \cm3) clouds are 
necessary to reproduce the critical IR line ratios observed in the 
low-excitation Starburst Nucleus Galaxies (SBNGs: M82, M83, NGC 253, NGC 3256, 
NGC 3690, and NGC 4945). These model predictions are in good agreement with 
the powerful starburst-driven superwinds and highly pressured ISM observed 
in SBNGs. On the contrary, the high-excitation \hii\ galaxies 
(II Zw 40 and NGC 5253) do not show any clear signature of large scale 
outflows of gas. This difference between \hii\ galaxies and SBNGs can 
be interpreted in terms of temporal evolution of their starbursts. 

%\keywords{galaxies: individual: NGC 3690; NGC 3256; M82; 
%M83; NGC 253; NGC 4945; NGC 5253; II Zw 40 -- galaxies: ISM -- 
%galaxies: starburst -- infrared: galaxies - line: formation}


\end{abstract}

%%%%%%%%%%%%%%%%%%%%%%%%%%%%%%%%%%%%%%%%%%%%%%%%%%%%%%%%%%%%%%%%%%%%%%%%%

\section{Introduction}

Detection of faint emission in the high excitation [\oiv] 25.9$\mu$m 
infrared (IR) line has been reported by Lutz et al.\ \cite*{LUTZetal98} 
in a number 
of well-known starburst galaxies from observations with the 
Short Wavelength Spectrometer on board {\sl ISO}.

IR line ratios such as [\oiv] 25.9$\mu$m / [\neii] 12.8$\mu$m and 
[\oiv] 25.9$\mu$m / [\neiii] 15.6$\mu$m were used 
previously \cite{LUTZetal96,GENZELetal98} to establish the dominant 
source (starburst or AGN) of ionizing radiations in ultraluminous infrared 
galaxies. 
Actually, rather than normalizing just to [\neii] or [\neiii], 
Lutz et al.\ \cite*{LUTZetal98} use [\oiv]/([\neii] + 0.44[\neiii]) 
as a critical line ratio to reveal different excitation conditions in starburst 
galaxies. 
Results from pure photoionization  models or from pure shock models are 
used by these authors in order to explain the observed line ratios. 
However, these models were not successful in reproducing the
IR line ratios of several starburst galaxies. In fact, models with a 
power-law radiation producing enough [\oiv] line give  too high  values
for the [\neiii]/[\neii] line ratio. 
Those with two-temperature (39000 K + 80000 K) black body radiation 
fail in explaining the low-excitation starbursts. 
Finally, pure shock models could provide [\oiv] line intensities similar to 
observations but not enough [\neii] and [\neiii] which should come from the 
\hii\ region.
 
These results indicate that  a composite model, 
which accounts for the coupled effects of shocks and
photoionization, must be used.  
Indeed, in a disturbed region such as that containing 
starbursts, it is expected that shocked gas is embedded in 
stellar radiation.
This kind of composite model has successfully explained the AGN 
spectral features (Viegas \& Contini 1997 and references 
therein) as well as those of single galaxies where
the AGN and starburst activities are of comparable importance 
(e.g. the Circinus galaxy, Contini et al.\ 1998).
These models were provided by the SUMA code 
(see Viegas \& Contini 1994  and references therein). 

In the present study we suggest that shocks result from supernova 
explosions in the nuclear 
starburst region of the galaxy, which provide enough kinetic energy 
to create high velocity superwinds. 
These in turn will accelerate gas clouds on large scales.
The physical conditions in the ionized interstellar clouds are determined 
by shocks coupled to photoionization by massive hot stars (O and Wolf-Rayet).

%TC
This paper, rather than being an answer to 
Lutz et al.\ \cite*{LUTZetal98}, is aimed at carrying on
their investigation. The distinction between AGN and starburst activity 
in galaxies being already established \cite{LUTZetal96,GENZELetal98}, 
our aim is to find out which models (provided by the SUMA code), 
better fit the observed trend between IR line ratios in starburst galaxies. 
%The code SUMA is used to model the IR lines of the observed starburst 
%galaxies. 
We will not discuss in detail the 
single galaxy structures, but  will use the fit of calculated
to observed line ratios to define the critical parameters
(and their ranges of values) which prevail in the emitting gas.
In other words, we will treat the 
emission-line regions of the observed galaxies 
as schematic systems of gaseous clouds moving radially outward from 
star-forming regions ionized by hot stars and shocks, avoiding  a deep 
investigation on the internal structure of single objects.

\begin{table}
\caption[]{Parameters for single-cloud models}
\begin{flushleft}
%{\footnotesize
\begin{tabular}{lrrrr}
\hline
\hline
Galaxy & \Vs & \n0 & $U$ & $D$ \\
       & (\kms) & (\cm3) & & (pc) \\
\hline
II Zw 40 & 50  & 100 &0.02 &0.10 \\
NGC 5253 & 100 & 100 &0.01 &0.01 \\
 & & & & \\
NGC 3690 B/C & 300 & 100 &0.01 &0.03 \\
NGC 3256 & 400 & 500 &0.01 &1.70 \\
M 82 & 400 & 500 &0.01 &2.70 \\
M 83 & 500 & 500 &0.01 &0.70 \\
NGC 253 &600 & 800 &0.01 &0.23 \\
NGC 4945 & 600 & 800 &0.01 &0.20 \\
\hline
\hline
\end{tabular}
%}
\end{flushleft}
\label{table1}
\end{table}

\section{The results by a composite model}
\label{model}

As commented above, Lutz et al.\ \cite*{LUTZetal98} 
concluded that shocks are the
most plausible ionizing mechanism for explaining the [\oiv] line, while
the [\neii] and [\neiii] come from the photoionized region. However,
shock results  should be obtained by consistent models
concomitantly with photoionization from hot stars.
%TC
The SUMA code consistently accounts for both the shock effect accompanying 
cloud expansion and photoionization by hot stars, in a plane-parallel 
symmetry. The geometrical thickness of the clouds range from 0.01 to 
2.70 pc (see Table~\ref{table1}), which is small compared with the 
radius of the starburst regions in galaxies.
In our model, we assume that the gas clouds move radially outward from 
the starburst region, i.e. the hot star population is facing the inner 
edges of the  clouds while a shock develops on the opposite 
side. 

For all models an average  $stellar \, temperature$, 
\Ts = $5 \times 10^4$ K, for the hot star population, 
and a preshock magnetic field, \B0 = $10^{-4}$ gauss, are adopted, 
as well as cosmic abundances.
Lower abundances should lead to a lower cooling rate of the gas 
downstream which could be reajusted by adopting  higher preshock densities.
%TC
The other input parameters for SUMA: the shock velocity, \Vs,
the preshock density, \n0, the ionization parameter\footnote{The 
ionization parameter is the ratio of the photon density of the ionizing 
radiations reaching the nebula to the gas density in the inner part of 
the nebula.}, $U$, and the
geometrical thickness of the clouds, $D$, are chosen to vary
within the ranges consistent with observations. 
%TC

\begin{table*}
\caption{Calculated line fluxes (in \erg) with multi-cloud models}
\begin{flushleft}
\tiny{
\begin{tabular}{|l|rrrrr|rrrr|rrrr|}
\hline
 &&& &&& &&& &&&& \\
{\normalsize Line}  & \multicolumn{5}{c|}{\normalsize M 82} & \multicolumn{4}{c|}{\normalsize NGC 3256} & 
\multicolumn{4}{c|}{\normalsize NGC 253} \\
 &&& &&& &&& &&&& \\
%\cline{2-5}
%\cline{7-9}
%\cline{11-13}
 &&& &&& &&& &&&& \\
($\mu$m) & A1 & A2 & A3 & A4 & Av & B1 & B2 & B3 & Av & C1 & C2 & C3 & Av \\
 &&& &&& &&& &&&& \\
\hline
\Cii\ 158&0.05 & 0.026 & 0.007 & 0.420& 2.56 & 0.03 & 0.012 & 0.007 & 1.6 & 0.03 & 0.07 & $6\times 10^{-4}$&7.06 \\ 
&0.01  & 16.5  & 27.0  & 56.5& -    & 0.04  & 75   & 24.5  &-    & 0.02 & 99.1 & 0.85 & -   \\ 
\Oiii\ 88&9.6 & 0.002 & 0.008 & 0.001 & 0.92 & 6.2 & 0.002 & 0.008 & 0.8 & 2.10 & $3\times 10^{-5}$ &0.0076& 0.78 \\
&6.30 & 4.2    & 89.0   & 0.39  &   -  & 13.9 & 23.8  & 62.4  &  -   & 11.3& 0.3 &88.5  &-     \\
\Oi\ 63 & 0.004&0.067 & 0.007 & 0.348 & 2.96 & 0.002 & 0.025 & 0.007 & 2.9 & 0.78 & 0.075 & $5\times 10^{-4}$ & 7.59 \\
& 0.0008 &37.0   & 23.0   & 40.0  &  -   & 0.002 &86.5   & 13.5  &  -  & 0.5 & 98.8  & 0.66  & -    \\
\Oiii\ 52 & 89.6 & 0.004 & 0.008 & 0.003 & 1.46 & 56.6 & 0.004 & 0.008 & 2.0 & 20 & $6.7\times 10^{-5}$ & 0.0078&1.73 \\
&36.7  & 4.5   & 57.8   & 0.9  & -  & 54.6 & 20.4  & 26.5  &  -  & 53.7& 0.4     &45.7   &-    \\
\Slii\ 35 & 25.6 & 0.166 & 0.027 & 2.170 & 13.20 & 15.9 & 0.073 & 0.027 & 9.4& 3.70 & 0.44 &0.01 & 45.2 \\
& 1.2  & 20.8  & 20.5  & 57.6 &  -   & 3.2  & 77.4  & 19.4  & -  & 0.4& 97.4 &2.2  & -    \\
\Siii\ 33 & 12.0 & 0.055 & 0.001 & 0.570 & 3.10 & 7.5 & 0.024 & 0.001 & 2.6 & 3.30 & 0.02 & $2\times 10^{-4}$ &2.17 \\
 & 2.3 & 29.3  & 4.2    & 64  &  -  & 5.5  & 91.6  & 3.0    &  -   & 7.1 & 91.9 & 0.92  & -   \\
\Siii\ 18 & 96.0 & 0.031 & $6.6\times 10^{-4}$ & 0.114 & 1.54 & 60.6 & 0.013 & $6.6\times 10^{-4}$ & 2.5 & 30 & 0.005& $1.2\times 10^{-5}$ &1.85 \\
& 1.9 & 32.9  & 4.2     & 25.8 & - & 46.2 & 52.2  & 0.02   &  -   & 75 & 24.5  & 0.64   & -   \\
 &&& &&& &&& &&&& \\
\Oiv\ 25.9 & 7 & $1.2\times 10^{-4}$ & $6.6\times 10^{-5}$ & $2.3\times 10^{-4}$ & 0.05 & 7  &$1.2\times 10^{-4}$ & 
$6.6\times 10^{-5}$ & 0.14 & 5.6 & $8.8\times 10^{-5}$ & $6.3\times 10^{-5}$&0.3 \\
& 82 & 4 & 13  & 1 &  -& 89   &8   & 3  & -  & 94& 3    &2    & -   \\
\Neiii\ 15.5&123 & 0.014 & 0.008 & 0.01 & 1.80 & 80 & 0.0075 & 0.008 & 2.77 & 46 & $8\times 10^{-4}$& 0.0063&2.8 \\
&40&  13  & 45  & 2  & -   & 55 & 27   & 18   &  -   & 75 & 3  & 22  & -  \\
\Neii\ 12.8 & 960 & 0.025 & 0.004& 0.15 & 7.06 & 606 & 0.01 & 0.004 & 12.6 & 490 & 0.0055 & 0.0016 & 23.3 \\
& 82 & 6 & 5 & 7& - & 90 & 8 & 2 & - & 97 & 2 & 0.7 & - \\
 &&& &&& &&& &&&& \\
$W$ (\%) & 0.005& 13.8&83.3 &2.895 &-&0.01 & 60.4 & 39.59&- & 0.007 &49.493 & 50.5&- \\
%P(\Cii\ 158)&0.01  & 16.5  & 27.0  & 56.5& -    & 0.04  & 75.   & 24.5  &-    & 0.02 & 99.1 & 0.85 & -   \\ 
%P(\Oiii\ 88)&6.30 & 4.2    & 89.0   & 0.39  &   -  & 13.9 & 23.8  & 62.4  &  -   & 11.3& 0.3 &88.5  &-     \\
%P(\Oi\ 63) & 0.0008 &37.0   & 23.0   & 40.0  &  -   & 0.002 &86.5   & 13.5  &  -  & 0.5 & 98.8  & 0.66  & -    \\
%P(\Oiii\ 52) &36.7  & 4.5   & 57.8   & 0.9  & -  & 54.6 & 20.4  & 26.5  &  -  & 53.7& 0.4     &45.7   &-    \\
%P(\Slii\ 35) & 1.2  & 20.8  & 20.5  & 57.6 &  -   & 3.2  & 77.4  & 19.4  & -  & 0.4& 97.4 &2.2  & -    \\
%P(\Siii\ 33) & 2.3 & 29.3  & 4.2    & 64.  &  -  & 5.5  & 91.6  & 3.0    &  -   & 7.1 & 91.9 & 0.92  & -   \\
%P(\Siii\ 18) & 1.9 & 32.9  & 4.2     & 25.8 & - & 46.2 & 52.2  & 0.02   &  -   & 75. & 24.5  & 0.64   & -   \\
% &&& &&& &&& &&&& \\
%P(\Oiv\ 25.9) & 82 & 4 & 13  & 1 &  -& 89   &8   & 3  & -  & 94& 3    &2    & -   \\
%P(\Neiii\ 15.5)&40&  13  & 45  & 2  & -   & 55 & 27   & 18   &  -   & 75 & 3  & 22  & -  \\
%P(\Neii\ 12.8) & 82 & 6 & 5 & 7& - & 90 & 8 & 2 & - & 97 & 2 & 0.7 & - \\
%P($\rm H\beta$)& 90 & 2 & 6 & 2&- & 95 & 4    & 2   &-& 97 &2  & 1 & - \\
\hline
\Vs\ (\kms)& 400 & 200 & 100 & 200 &-& 400 & 200 & 100 &-& 600 & 100 & 100 &- \\
\n0\ (\cm3)& 500 & 50 & 40 & 100 &-& 500 & 50 & 40 & -& 800 & 200 & 40 & - \\
$U$ & 0.01 & 0.01 & 0.01 & 0.01 &- & 0.01 & 0.01 & 0.01 &-& 0.01 & 0.001 & 0.01&- \\
$D$ (pc) & 2.67 & 3.33 & 3.33 & 20&-& 1.67 & 1.67 & 3.33 &-& 0.23 & 10 & 0.5& - \\
\hline
\noalign{\smallskip}
\noalign{\small For each line, we report as a second entry the percentage 
contribution of each model in the weighted sum flux.}
\end{tabular}
}
\end{flushleft}
\label{table2}
\end{table*}

\subsection{Single-cloud models}
\label{scm}
%{\it Single-cloud models --}

\begin{figure}[t]
  \resizebox{\hsize}{!}{\includegraphics{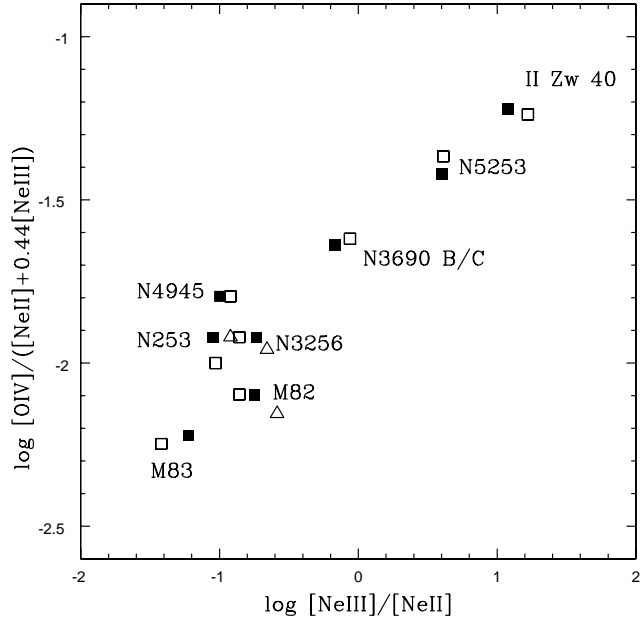}}
  \caption{
[\oiv] 25.9$\mu$m / ([\neii] 12.8$\mu$m + 0.44 [\neiii] 15.5$\mu$m) 
versus the [\neiii] 15.5$\mu$m / [\neii] 12.8$\mu$m line ratio. 
The observational data from Lutz et al.\ (1998) 
are indicated by filled squares. Open symbols correspond to results 
for single-cloud (squares, input parameters listed in 
Table~\ref{table1}) and multiple-cloud (triangles, input 
parameters listed in Table~\ref{table2}) models. 
  }
  \label{fig1}
\end{figure}

First we perform calculations by single-cloud models, 
which are compared with the observations of each galaxy in Fig.~\ref{fig1}. 
The input parameters in the models 
%and the observed global properties which 
%characterize each galaxy 
are given in Table~\ref{table1}.
Filled squares represent the data from the 
Lutz et al.\ \cite*{LUTZetal98} 
sample  and open squares the results of model calculations.
Upper limits are not included. As shown in 
Lutz et al.\ (1998, Fig. 2) they are located in 
the left region of Fig.~\ref{fig1}. 
 The fit is very good for almost all galaxies. 
A perfect fit for individual data is senseless because many different 
conditions coexist in single objects.   
Notice indeed that NGC 4945 is a composite object with  
starburst/Seyfert 2 nuclear activity. This could explain its higher 
intensity of [\oiv] 25.9$\mu$m relative to model predictions for 
starburst only. A possible hidden AGN has also been reported in the 
merger galaxy NGC 3256 \cite{KOTILAINENetal96}.

The parameters listed in Table~\ref{table1} indicate that the 
shock velocity and the preshock density are the critical parameters 
to reproduce the observed 
trend of the IR line ratios in Fig.~\ref{fig1}.  
The velocity growth is generally followed by an increase of the preshock
density. 
An ionization parameter of 0.01 was adopted for nearly all the models 
except for II Zw 40 for which $U = 0.02$.
For this  galaxy the effect of radiation dominates the shock effect 
(see Sect.~\ref{obs}). The geometrical thickness of the clouds, $D$, 
ranges between 0.01 pc for NGC 5253 and $<$ 3 pc for M 82.

Lutz et al.\ \cite*{LUTZetal98} already suggested that shocks 
are the most plausible source of [\oiv] emission observed in starburst 
galaxies. Our results  
show that shocks are also necessary to reproduce the low observed values 
of [\neiii]/[\neii] (see Sect.~\ref{physic} for details). 
Moreover, Fig.~\ref{fig1} shows that the objects with [\neiii]/ [\neii] 
$<$ 1 
are well fitted by models characterized by high velocity and density clouds. 
However, previous investigations on the physical conditions in the ISM of 
NGC 253 \cite{CARRALetal94}, 
NGC 3256 \cite{CARRALetal94,RIGOPOULOUetal96}, 
and M 82 \cite{HOUCKetal84,LORDetal96} 
found relatively low electronic densities on the basis of the 
[\siii] 33$\mu$m/18$\mu$m and [\oiii] 88$\mu$m/52$\mu$m line ratios, 
and low velocities derived from the FWHM of the IR line profiles.

\begin{table}
\caption[]{Observed vs. calculated IR line fluxes$^{\rm a}$}
\begin{flushleft}
{\footnotesize
\begin{tabular}{lrrrrrr}
\hline
\hline
Line & \multicolumn{2}{c}{NGC 253} & \multicolumn{2}{c}{NGC 3256} & \multicolumn{2}{c}{M 82} \\
($\mu$m) & Obs$^{\rm b}$ & Cal & Obs$^{\rm b}$ & Cal & Obs$^{\rm b}$ & Cal \\
\hline
\Oiii 88 & 0.10$\pm$0.02  & 0.1 & 0.4$\pm$0.3 & 0.5 & 0.6$\pm$0.2 & 0.4 \\
\Oi 63 & 0.9$\pm$0.1  & 1.0 & 1.2$\pm$0.4 & 1.7 & 1.1$\pm$0.4 & 1.2 \\
\Oiii 52 & 0.20$\pm$0.04  & 0.2 & $<$1.3 & 1.2 & 0.7$\pm$0.2 & 0.5 \\
\Slii 35 & 1.1$\pm$0.2 & 6.2 & 1.9$\pm$0.5 & 5.6 & 1.0$\pm$0.4 & 5.4 \\
\Siii 33 & $<$1.3  & 0.3 & 1.2$\pm$0.3 & 1.6 & 1.1$\pm$0.3 & 1.2 \\
\Siii 18 & \ldots & 0.2 & 0.9$\pm$0.2 & 1.5 & 0.7$\pm$0.3 & 0.6 \\
\hline
\hline
\noalign{\smallskip}
\noalign{\small $^{\rm a}$ Fluxes are relative to [C\, {\sc ii}] 158$\mu$m.}
\noalign{\small $^{\rm b}$ Observed line fluxes from Carral et al.\ 
\cite*{CARRALetal94} for NGC 253 and NGC 3256, 
Rigopoulou et al.\ \cite*{RIGOPOULOUetal96} for NGC 3256, 
and Lord et al.\ \cite*{LORDetal96} for M82.}
\end{tabular}
}
\end{flushleft}
\label{table3}
\end{table}

\subsection{Multi-cloud models}
%{\it Multi-cloud models --}

Many different conditions coexist in the ISM of single galaxies. 
Therefore, we performed multi-cloud model calculations to reproduce all the 
observed IR line ratios available in the literature for the starburst galaxies 
NGC 253, NGC 3256, and M 82. 
In these models, we assume that both {\it low} and {\it high} 
density-velocity (d-v) clouds are present in single objects.

In Table~\ref{table2}, the calculated IR line fluxes are given for 
single-cloud models and for the weighted sums of the multi-cloud models. 
Notice that the line fluxes are calculated at the 
emitting nebula while the observations are given at Earth. 
The input parameters of single-cloud models are also listed.

Line fluxes are proportional to the square of the gas emitting density, and 
compression downstream increases with the shock velocity. Thus, 
the line fluxes computed from high d-v models (i.e. A1, B1, and C1) are higher 
than those calculated from low d-v models (A2, A3, A4, B2, B3, C2, and C3) 
by some orders of magnitude, except for lines \Oi\ and \Cii.
The gas in high d-v models is ionized to very high temperatures 
(Fig.~\ref{fig2}b), and so low-level lines like \Oi\ and \Cii\ are weak. 
Recall that the first ionization potential of C is lower than
that of H (and Ne). Consequently, \Cii\ lines, as well as \Oi\ lines,
are emitted by relatively cold gas and can be weak in models
which show strong \Neii\ lines. 

%TC
In order to fit the observations, a balance must be achieved between high 
and low d-v models in the multi-cloud spectra, and the final result is a 
weighted sum (A$_{\rm v}$ in Table 2) with the weights chosen adequately. 
For each model, the weight 
$W$ represents the relative number of clouds (in \%) characterized by that 
model. The weights of low d-v models must be by a factor of $10^4 - 10^5$ 
higher than for high d-v models. 
In other words, the number of high d-v clouds must be very low compared with 
the number of low d-v clouds.
In Table~\ref{table2}, for each emission line, the second entry 
corresponds to the percentage contribution of the single-cloud model to 
the line flux, taking into account the corresponding weight. 
The high d-v clouds 
contribute up to 90\% to the \Oiv\ and \Neii\ emission lines, 
whereas the \Neiii\ line is produced at a same level both by low and high 
d-v clouds for M 82 and NGC 3256. 
The others IR lines are mainly emitted by the low d-v clouds except 
the \Oiii\ 52 line. 

The results of multi-cloud model calculations are compared to 
the observed IR line ratios in Table~\ref{table3} and plotted in 
Fig.~\ref{fig1} (open triangles). 
The fit is good for all line ratios (especially for the density critical 
ones, see sect.~\ref{scm}), except for the [\slii] line which is 
overpredicted by a factor of about 5. 
This might indicate that silicium is depleted and locked in 
dust grains. Adopting a relative abundance Si/H lower by a factor of $\sim$ 5 
than cosmic will not affect the predictions for the other line ratios 
listed in Table~\ref{table3}. 
Most of the IR lines reported in Table~\ref{table2} 
(e.g. [\oi], [\cii], etc) 
are mainly produced by the low d-v clouds.
Thus, in the weighted sum, the IR lines ratios [\siii] 33$\mu$m/18$\mu$m 
and [\oiii] 88$\mu$m/52$\mu$m 
are determined by low d-v models, while the [\oiv]/([\neii]+0.44[\neiii])
and [\neiii]/[\neii] line ratios are mainly determined by the high d-v cloud 
component.

\begin{figure}[t]
  \vspace{-1cm}
  \resizebox{\hsize}{!}{\includegraphics{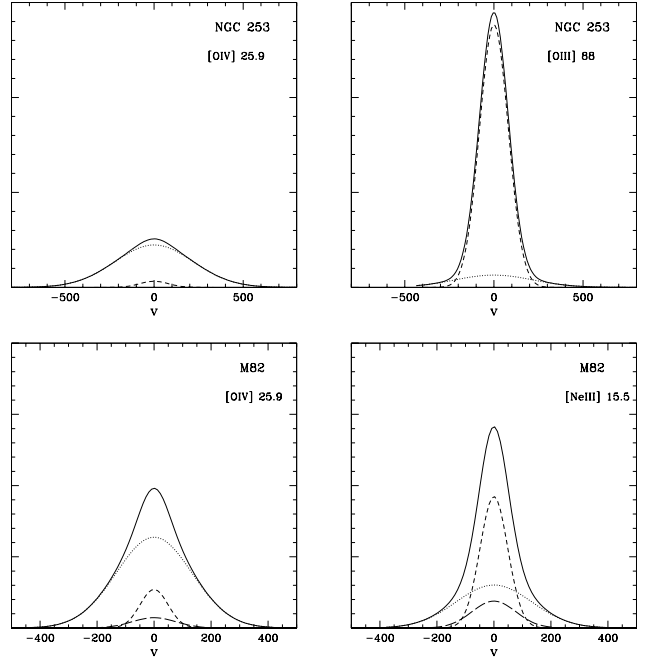}}
  \vspace{-2.5cm}
  \caption{Predicted line profiles for NGC 253 (top) and M 82 (bottom) using 
multi-cloud models. The final profile (solid line) is the weighted sum 
of different cloud components: high density-velocity clouds (dotted line) 
and low density-velocity clouds (long-dashed line: \Vs\ = 200 \kms; 
short-dashed line: \Vs\ = 100 \kms)}
  \label{lineprof}
\end{figure}

\begin{figure*}[]
a)
\psfig{figure=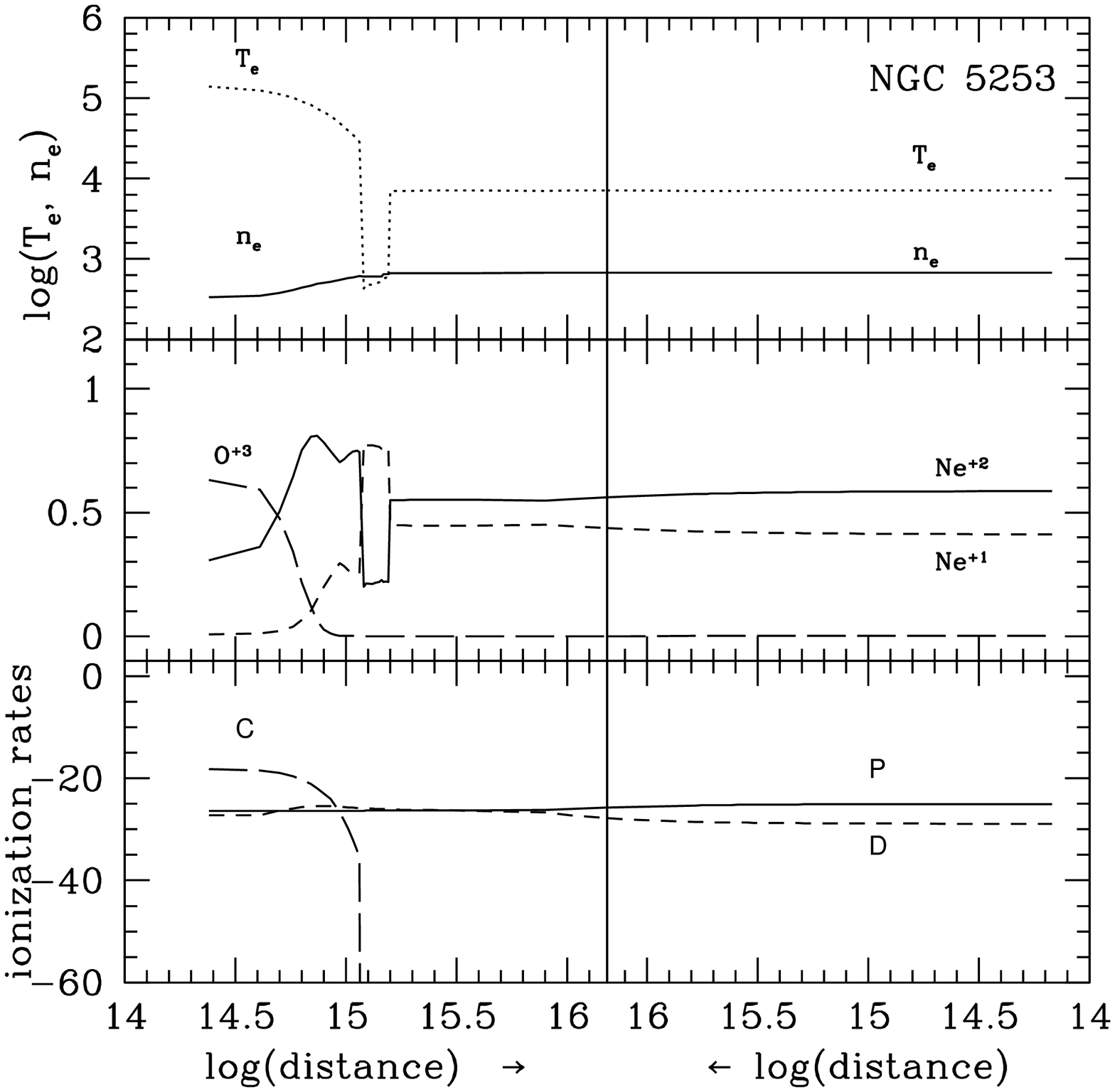,height=8.8cm}
b)
\psfig{figure=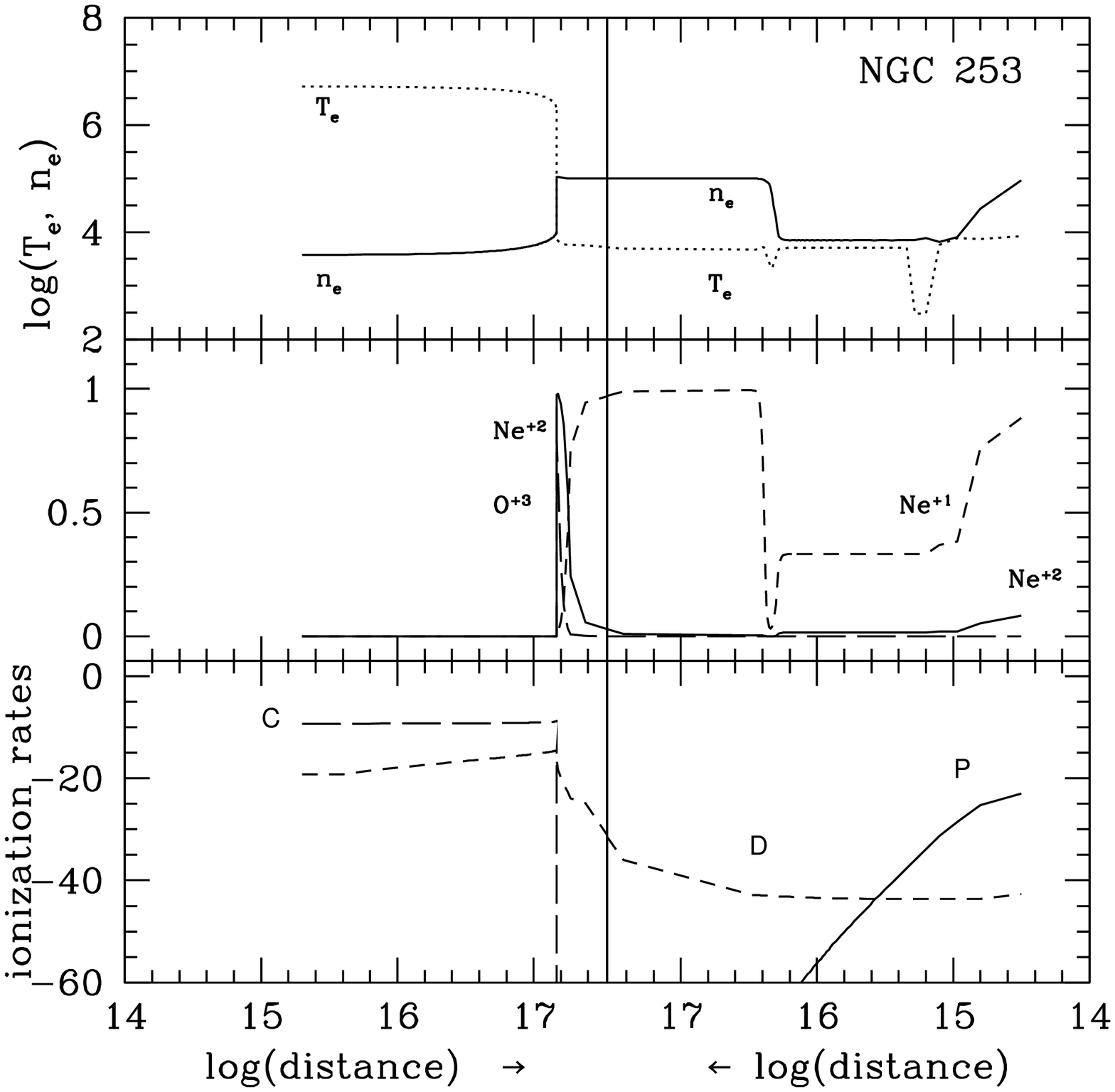,height=8.8cm}
\caption{
The physical conditions across low-density (a, NGC 5253) and high-density 
(b, NGC 253) clouds are shown in three panels, each one with a vertical 
line indicating the center of the emitting clouds. 
The electron temperature (dotted line) and electron density (solid line) 
appear in the top panel. 
The fractional abundances of $\rm O^{+3}$ (long-dashed line), 
$\rm Ne^{+2}$ (solid line), and $\rm Ne^{+1}$ (short-dashed line) are 
displayed in the central panel. 
The ionization rates for $\rm Ne^{+2}$ (in logarithmic units) due to 
electron collisions C (long-dashed line), to the stellar radiation P 
(solid line) and to the diffuse radiation D (short-dashed line) are 
shown in the bottom panel.  
The shock front is on the left side and the edge of the cloud 
photoionized by the stellar radiation is on the right of each panel. 
The horizontal axis scale is logarithmic and symmetric in order to show 
an equal view 
of both sides of the emitting cloud
}
\label{fig2}
\end{figure*}

Once fitted the line ratios, it is now possible to examine the
FWHM of the line profiles, because the weights adopted in Table~\ref{table2} 
for high d-v models are very low and could lead to hardly observable broad 
components in the line profiles. 
The calculated line profiles for M 82 and NGC 253 are 
shown in Fig~\ref{lineprof}. The effect of galaxy rotation is included 
in the calculation. Rotation velocities are about 76 \kms\ and 189 \kms\ for 
M 82 and NGC 253 respectively.
Two significant lines are chosen for each galaxy: \Oiv\ and \Neiii\ for M 82, 
and \Oiv\ and \Oiii\ 88 for NGC 253. The \Oiii\ 88 line is chosen because it 
represents the case of a low contribution of the high d-v model (C1) 
relatively to a strong contribution of the low d-v model (C3). 
Single-cloud model profiles are also plotted in the figures using their 
percentage contribution reported in Table~\ref{table2}.
The broad component from the high d-v clouds dominates the weighted averaged 
profile of \Oiv, while in the \Oiii\ 88 and \Neiii\ profiles 
the broad component is faint and hardly observable, due to the low 
contribution from high d-v clouds. For M 82, the predicted FWHM of the 
\Oiv\ profile is $\sim$ 300 \kms, in good agreement with the observed 
value (FWHM $\sim$ 360 \kms) reported by Lutz et al.\ \cite*{LUTZetal98}. 

\section{The physical conditions across emitting clouds}
\label{physic}

To better understand the role of the input parameters in determining 
the physical conditions of the emitting gas, the distribution of some 
important physical quantities across the cloud is given in 
Figures~\ref{fig2}a and \ref{fig2}b  respectively for two significant 
models listed in Table~\ref{table1}: NGC 5253, which represents the 
low density-velocity cloud component, and the high density-velocity 
component of NGC 253. Each figure shows three 
panels where the shock front is on the left and the photoionized edge 
is on the right. The top panel displays the electron temperature, \Te, 
and density, \ne. The fractional abundances of the ions $\rm O^{+3}$/O, 
$\rm Ne^{+2}$/Ne, and $\rm Ne^{+1}$/Ne are shown in the central panel.  
The ionization rates for $\rm Ne^{+2}$ due to the primary radiation 
(P), to the diffuse radiation (D), and to electronic collisions (C) 
are given in the bottom panel. 

The  [\oiv] line is always 
emitted from the shock dominated region where the temperatures are high,
with no contribution from the photoionized zone. 
In particular, Fig.~\ref{fig2}a  shows that the  critical temperatures
are about $10^5$ K.
For NGC 5253 the shock effect becomes negligible at a 
distance of about $10^{15}$ cm from the shock front, and most of the 
physical conditions in the cloud are then dominated by the effect of 
stellar radiation which leads to [\neiii]/[\neii] $>$ 1.
On the other hand, due to strong compression in the NGC 253 gas clouds 
(Fig.~\ref{fig2}b), the temperature rapidly  drops to \Te\ $< 10^4$ K.
This can be explained by the high density ($\sim 10^5$ \cm3) of the gas
in the innermost region of the cloud which strengthens the cooling rate.
The role of the primary radiation is to keep the gas at a temperature 
slightly below $10^4$ K corresponding to a relatively high $\rm Ne^{+1}$/Ne 
compared to $\rm Ne^{+0}$/Ne and $\rm Ne^{+2}$/Ne. 
Therefore, for this galaxy, [\neiii]/[\neii] is  $<$ 1.
Owing to the high optical thickness of the gas, the stellar radiation 
flux reaching the inner edge of the cloud rapidly decreases, 
the size of the $\rm Ne^{+2}$ photoionized zone is reduced, and the diffuse 
radiation effect prevails accross the cloud. 

For NGC 253, temperatures as high as $7 \times 10^6$ K deduced 
from the ROSAT PSPC data (Heckman 1996 and references
therein) are in agreement with $5.4 \times 10^6$ K obtained by the 
model in the postshock region (Fig.~\ref{fig2}b). 
%The observed X-ray luminosity in the ROSAT band of $4 \times 10^{39}$ 
%\ergs\ can be compared to a mechanical luminosity of the starburst 
%of $\sim 10^{42}$ \ergs\ (Heckman 1996 and references 
%therein). This value is in agreement with the calculated X-ray flux 
%(20 \erg) if the emitting surface has a radius of about 40 pc. 
%Indeed, if a filling factor is taken into account a higher radius is 
%required. 
Precursor soft X-rays radiation from the hot gas could affect the 
ionization conditions in the upstream gas clouds. However, due to the 
adiabatic jump and compression, the densities $n$ upstream are lower 
than in the downstream gas by at least a factor of 4.
Considering that line intensities depend on $n^2$, the contribution of 
the line emission from upstream gas to the final spectrum will be small.

\section{Comparison with the observed starburst properties}
\label{obs}

One might interpret the observed trend in Fig.~\ref{fig1} as a 
continuous sequence between two extreme cases: 
starburst regions where shocks in high-velocity clouds are 
{\em necessary} to reproduce the observed critical IR line ratios 
(bottom left part of the diagram) and the ``radiation-dominated'' ones 
(upper right part of the diagram), where the shock contribution is 
relatively less important, because formed only in low-velocity clouds.
Let us now check wether the critical parameters derived from the models 
along this sequence are in agreement with the observed properties 
of the starburst galaxies.

\begin{table}
\caption[]{Galaxian properties}
\begin{flushleft}
%{\footnotesize
\begin{tabular}{llrrr}
\hline
\hline
Galaxy & Morph.$^{\rm a}$ & $M_{\rm abs}^{\rm a}$ & $d^{\rm b}$ & $\log L_{\rm X}^{\rm b}$ \\
       &        &               &(Mpc)& (\ergs) \\
\hline
II Zw 40 & BCDG & $-$14.5 & 9.2 & ... \\
NGC 5253 & Irr & $-$16.2 & 4.1 & 38.6 \\
 & & \\
NGC 3690 B/C & Merger & $-$21.1 & 42.4 & 41.0 \\
NGC 3256 & Merger & $-$22.2 & 33.2 & 41.7  \\
M 82 & Sd & $-$19.0 & 4.3 & 39.7 \\
M 83 & SABc & $-$21.2 & 5.1 & 40.2 \\
NGC 253 & SABc & $-$20.9 & 3.3 & 39.9 \\
NGC 4945 & SBcd & $-$20.1 & 5.1 & 39.9 \\
\hline
\hline
\noalign{\smallskip}
\noalign{\small $^{\rm a}$ From the LEDA database}
\noalign{\small $^{\rm b}$ The 
distance ($d$) is derived from the Galactic Standard of Rest (GSR) 
velocity, using a Hubble constant $H_0 = 75$ km s$^{-1}$ Mpc$^{-1}$, 
except for the distance of NGC 5253 taken from Saha et al.\ (1995). 
The X-ray luminosity is derived from the flux reported in the 
{\em ROSAT} catalog (Brinkmann et al.\ 1994) or in 
the {\em EINSTEIN} catalog (Fabbiano et al.\ 1992).}
\end{tabular}
%}
\end{flushleft}
\label{table4}
\end{table}

%TC
Galaxies in the first group (the last six galaxies in Table~\ref{table4}) 
are massive 
and chemically evolved spiral galaxies which exhibit a single 
energetic and compact starburst in their nucleus, and are commonly 
named Starburst Nucleus Galaxies (SBNGs; Balzano 1983; 
Coziol et al.\ 1994). 
Evidence for powerful large-scale starburst-driven superwinds in the nuclear 
region of SBNGs 
(like M 82, NGC 253, NGC 4945 and NGC 3690 in the present sample) 
has been reported by Heckman et al.\ \cite*{HAM90}. 
Moreover, the strong X-ray emission 
($L_{\rm X} \sim 10^{40}-10^{42}$, see Table~\ref{table4}) 
in SBNGs provides strong evidence 
for the formation of hot superbubbles due to the combined action 
of stellar winds from massive stars and supernova 
explosions in the central starburst cluster.   
Expanding velocities as high as $\sim 500 - 600$ \kms\ have been 
estimated for the ionized gas in 
M 82 \cite{SB-H98}, 
NGC 4945 \cite{MOORWOODetal96}, 
and NGC 253 \cite{H96}. 

In the central region of SBNGs, 
part of the ISM is overpressured relative to the ISM in the Milky Way 
\cite{HAM90,CARRALetal94}, 
due to compression by superwinds. 
The nuclear starbursts usually contain 
many compact radio sources associated with bright and small 
supernova remnants 
(Muxlow et al.\ 1994; 
Huang et al.\ 1994 for M 82; 
Ulvestad \& Antonucci 1997 for NGC 253; 
Zhao et al.\ 1997 for NGC 3690). 
Such a population of dense and compact objects is generally 
attributed to the high pressure in the central region of SBNGs which allows 
the formation of dense ionized clouds (see e.g. Carilli 1996).

The galaxies in the second group (the first two galaxies 
in Table~\ref{table4}) 
are mostly metal-poor dwarf irregular galaxies with multiple star-forming 
regions distributed randomly, and are known as \hii\ galaxies \cite{F80}.
The arche\-types of this class of galaxies are II Zw 40 and NGC 5253. Their
starbursts are very young ($\leq$ 4 Myr), as proved by the 
numerous Wolf-Rayet stars still present in II Zw 40 \cite{KS81,VC92} 
and NGC 5253 \cite{SCHAERERetal97,SCK99}, and only contain a few supernova
remnants (Vanzi et al.\ 1996 for II Zw 40; 
Beck et al.\ 1996 for NGC 5253). Moreover, 
there is  no evidence for powerful starburst-driven superwinds 
in these galaxies (Martin 1998 for II Zw 40; 
Marlowe et al.\ 1995 for NGC 5253) which are faint 
or even undetected in X-rays 
(see Table~\ref{table4}).

Thus, the distinction between \hii\ galaxies and SBNGs can be understood 
in terms of temporal evolution of the starbursts. 
In the former galaxies (II Zw 40 and NGC 5253), the star-forming event 
is still young enough to contain a high proportion of massive O and 
Wolf-Rayet stars. 
These populations of hot ($T_{\rm eff} \sim 60 000$ K) stars produce 
large amounts of ionizing radiation, but did not have the time to 
evolve off the main sequence to produce supernova explosions. 
In SBNGs, like M 82 or NGC 253, the starburst is much older 
($\geq$ 20 Myr, e.g. Engelbracht et al.\ 1998), 
and the most massive stars had the time to evolve 
through supernovae.
As shown by the models of Leitherer \& Heckman \cite*{LH95}, 
the ratio between the mechanical energy and the ionizing Lyman continuum, 
both injected in the ISM by the massive star population, rapidly increases 
by a factor of $\sim$ 100 between $\sim$ 5 and 20 Myr. 
This explain the difference between the ``radiation-dominated'' \hii\ 
galaxies and the ``shock-dominated'' SBNGs. 
The presence of high density-velocity clouds in SBNGs, which host a 
relatively ``old'' starburst, require large quantities of mechanical 
energy to power large-scale superwinds. These energetic events disrupt 
molecular clouds through shock waves and drive them out of the nucleus 
at high velocities. 

\section{Conclusions}
\label{conclu}

The observed relation between [\oiv]/([\neii]+ 0.44[\neiii]) 
and [\neiii]/[\neii], and single object IR line ratios are well 
reproduced by models accounting consistently for both shock and 
photoionization effects occurring in starburst regions. 
The critical parameters which reproduce the observed trend of the IR 
line ratios are the preshock density and the shock velocity which are 
consistently correlated.
Most of the shocks are produced in low density-velocity (\n0 $\sim$ 100 \cm3\ 
and \Vs\ $\sim 50 - 100$ \kms) clouds which represent the bulk of 
the ionized gas in starburst galaxies. 
However, even if they are by many orders less numerous, high-velocity 
($\sim 400 - 600$ \kms) shocks in dense ($\sim 500 - 800$ \cm3) clouds are 
necessary to reproduce the critical IR line ratios observed in the 
low-excitation SBNGs. 

These model predictions are in good agreement with the observed ISM 
properties in starburst galaxies. SBNGs are the only ones to exhibit 
powerful starburst-driven superwinds and highly pressured ISM. 
On the contrary, the high-excitation \hii\ galaxies do not show any 
clear signature of large scale outflows of gas. 
This difference between \hii\ galaxies and SBNGs can 
be interpreted in terms of temporal evolution of their starbursts. 
In the former, the starburst is too young to release large amounts 
of mechanical energy through supernova explosions whereas in the 
latter, characterized by older starbursts, the kinetic energy 
compete with the ionizing radiation in the heating of the ISM. 

\begin{acknowledgements}
M.C. is grateful to the Instituto Astron\^omico e Geof\' \i sico, USP 
for warm hospitality.
%We are grateful to an unknown referee for precious comments and 
%suggestions. 
We thank Emmanuel Davoust for a careful reading of the 
manuscript and the anonymous referee for helpful comments and suggestions. 
This research has made use of the Lyon-Meudon Extragalactic Database 
(LEDA) supplied by the LEDA team at the CRAL - Observatoire de Lyon (France). 
This work was partially supported by FAPESP and PRONEX/\-Finep, 
Brazil. 
\end{acknowledgements}

\end{document}